\documentclass[12pt]{article}
\usepackage{amsfonts,amssymb}
\newcommand{\End}{\nonumber\\}
\newcommand{\Bbc}[1]{{\mathbb #1}}
\newcommand{\Eps}{\epsilon}
\newcommand {\Cj}[1]{{#1}{}^*{}}
\newcommand{\Sum}[2]{\sum_{#1}^{#2}}
\newcommand{\Sumii}{\sum_{n=-\infty}^{+\infty}}
\newcommand{\Sumkii}{\sum_{k=-\infty}^{+\infty}}

\newcommand{\Dppp}[3]{\big\langle #1| #2 | #3 \big\rangle}
\newcommand{\Dpp}[2]{\big\langle #1| #2 \big\rangle}
\newcommand{\Ket}[1]{| #1 \big\rangle}
\newcommand{\Amom}{L}
\newcommand{\Apos}{\hat{x}}
\newcommand{\Lag}{{\cal L}}
\newcommand{\Real}{{\Bbc R}}
\newcommand{\Comp}{{\Bbc C}}
\newcommand{\Hil}{{\cal{H}}}
\newcommand{\Intg}{{\Bbc Z}}
\newcommand{\Trace}{{\rm Tr}}
\newcommand{\Hphi}{\hat{\psi}}
\newcommand{\Exp}[1]{\exp \left( #1 \right)}
\newcommand{\Dt}{\Delta t}
\newcommand{\Mod}{{\rm mod}}
\newcommand{\Factx}{ \frac{(i N)^{-p/2}}{\sqrt{ i a N p}}}
\newcommand{\Facty}{ \frac{(i N)^{-p/2}}{\sqrt{ i  N p}}}
\begin{document}
\begin{flushright}
 KCL-MTH-00-15
\end{flushright}
\begin{center}
 {\Large Gauss Sums and Quantum Mechanics}\\
 \ \\
 {\large Vernon Armitage\\
  Department of Mathematical Sciences\\
  University of Durham
  \footnote{Science Laboratories, South Road, Durham   DH1 3LE,
  J.V.Armitage@durham.ac.uk}\\
 \ \\
  and\\
  \ \\
  Alice Rogers\\
  Department of Mathematics\\
  King's College London
  \footnote{Strand,  London  WC2R 2LS,
  alice.rogers@kcl.ac.uk}}\\
 \  \\
  February 2000\\
 \  \\
\end{center}
%
\newtheorem{Def}{Definition}[section]
\newtheorem{The}{Theorem}[section]
\newtheorem{Lem}{Lemma}[section]
\bibliographystyle{unsrt}
\begin{abstract}
By adapting Feynman's sum over paths method to a quantum
mechanical system whose phase space is a torus, a new proof of the
Landsberg-Schaar identity for quadratic Gauss sums is given. In
contrast to existing non-elementary proofs, which use infinite
sums and a limiting process or contour integration, only finite
sums are involved. The toroidal nature of the classical phase
space leads to discrete position and momentum, and hence discrete
time. The corresponding `path integrals' are finite sums whose
normalisations are derived and which are shown to intertwine
cyclicity and discreteness to give a finite version of Kelvin's
method of images.
\end{abstract}
%
%
\section{Introduction}\label{INTRO}
In this paper we give a new proof, using only finite sums and
avoiding the need for analytic methods, of the Landsberg-Schaar
formula for quadratic Gauss sums. The key idea, which springs from
earlier work of one of us \cite{Armita87}, is to apply the
quantization of a quantum mechanical system whose phase space is a
torus, with both position and momentum (and also time) discrete
and cyclic. On adapting Feynman's `sum over paths' method to this
system a discrete and finite version of Kelvin's method of images
is obtained.

The formula states that for positive integers $p$ and $q$
\begin{equation}\label{LSeq}
  \frac{1}{\surd p} \sum_{n=0}^{p-1} \exp \left(  \frac{2\pi i n^2 q}{p} \right)
   = \frac{e^{\pi i/4}}{\surd(2q)}
   \sum_{n=0}^{2q-1} \exp \left( - \frac{\pi i n^2 p}{2q}\right).
\end{equation}
In number theory this formula plays a central r\^ole, underpinning
key results relating to quadratic reciprocity and characters. The
formula also promises to be a useful adjunct to discrete Fourier
transform methods used, for instance, in some algorithms for
quantum computing \cite{Shor94}. In this context it essentially
provides the discrete inverse quantum Legendre transformation from
the free Hamiltonian to the free Lagrangian.

The standard proof of the Landsberg-Schaar formula (as given, for
example, in \cite{DymMck}) is obtained by putting $\tau = 2iq/p +
\Eps$, $\Eps
> 0$, in the Jacobi identity for the theta function:
 \begin{equation}\label{JIeq}
   \sum_{n=-\infty}^{+\infty} e^{- \pi n^2 \tau}
   = \frac{1}{\surd \tau} \sum_{n=-\infty}^{+\infty} e^{-\pi n^2/\tau},
\end{equation}
and then letting $\Eps \to 0$. This method is an example of
Hecke's observation that `exact knowledge of the behaviour of an
analytic function in the neighbourhood of its singular points is a
source of arithmetic theorems' \cite{Hecke23} p225.

While this method certainly establishes the truth of the formula,
Hecke's insight notwithstanding it is somewhat unsatisfactory to
have to use analytic methods and take limits when only finite sums
are involved. The proof given in this paper does remain entirely
in the arena of finite sums; it demonstrates the extraordinary
range of Feynman's method of `summing over histories', which here
extends to the discrete and cyclic regime in a pleasing and
effective way. So, in a sense, the quantum mechanics offers
another example of Hecke's insight

The key ingredient in this paper is quantization on a toroidal
phase space.  A first step was taken by one of us in
\cite{Armita87}, leading to a proof of the Jacobi identity by
using random walks to quantize a system with cylindrical phase
space.  The key idea here is the observation that the  series
 $\vartheta (it) = \sum_{n=-\infty}^{+\infty} e^{- \pi n^2 it}$
is the trace of a certain quantum evolution operator
 $\exp -iHt/\hbar$, which converges in the distribution sense.
Using path integral methods adapted to the cylinder, in particular
Kelvin's method of images \cite{DymMck}, an  alternative formula
for the trace is found, and equating the two  expressions gives
the Jacobi identity. This work is described in  section
\ref{CPSsec} as an introduction to the main result in  this paper.

With a minor change of emphasis, the formula (\ref{LSeq}) turns up
by taking the operator that gives (\ref{JIeq}) (for $\tau = it$)
and then `looking at the system at discrete times $\tau = -2i/p$'.
This idea is pursued in section \ref{TPSsec}. In order to remove
difficulties over convergence, and to obtain the Landsberg-Schaar
formula (\ref{LSeq}), we change the problem to one in which
instead of the infinite set of energy levels (\ref{EVeq}) there is
only a finite number of energy levels $E_n$. This is achieved,
following a suggestion of Berry \cite{Berry1}, by replacing the
cylindrical phase space (with angular position $\theta$ periodic
but angular momentum $\Amom$ unbounded) by a torus on which both
$\theta$ and $\Amom$ are periodic. On quantization both $\theta$
and $\Amom$ are discrete, and moreover, as we see, discrete times
are also required. Adapting the methods used in section
\ref{CPSsec} to prove the Jacobi identity to this new setting,
again two expressions for the trace of the quantum evolution
operator are obtained, one by working in the momentum basis in
which the evolution operator is diagonal and the other by path
integral methods (using a novel discrete variant on Kelvin's
method of images); the equality of these two expressions gives the
Landsberg-Schaar identity.

This first description of the discrete and cyclic path integral
method is presented somewhat heuristically, but in  a manner which
should emphasise the ideas and motivation. In section
\ref{NORMsec} full details of the toroidal quantization are given
together with a justification of the normalisations used in
section \ref{TPSsec}.

\section{Cylindrical phase space and the Jacobi identity}
\label{CPSsec}

In this section earlier work of one of us \cite{Armita87} is
described as a prelude to the main result of the paper, in order
to demonstrate some of the novel features of the work.

Consider a rigid body constrained to rotate about a fixed axis (cf
Schulman \cite{Schulm68}). Let $I$ denote the moment of inertia
and $\theta$ denote the angle of rotation (given in  radians so
that it is taken mod $2\pi$).  An angular Schr\"odinger picture is
used, with the angular momentum observable $\Amom$ represented as
the operator
 \begin{equation}\label{EAMeq}
   \Amom    = - i\hbar \frac{ \partial}{\partial \theta}
 \end{equation}
(where $\hbar$ as usual stands for Planck's constant $h$ divided
by $2\pi$). Using the classical expression for the energy or
Hamiltonian
 \begin{equation}\label{EEeq}
  H = \frac{\Amom^2}{2I},
 \end{equation}
the Schr\"odinger equation for the wave function $\psi(\theta,t)$
is then \cite{Schiff}
 \begin{equation}\label{SEeq}
  i \hbar \frac{\partial \psi}{\partial t} = - \frac{\hbar^2}{2I}
  \frac{\partial^2 \psi}{\partial \theta^2}.
 \end{equation}
Suppose that $E_n$ denotes the $n^{th}$ eigenvalue of the
quantized Hamiltonian with energy eigenfunction $\phi_n$,
 \begin{equation}\label{EEEeq}
  -\frac{\hbar^2}{2 I} \frac{\partial^2}{\partial \theta^2}
  \phi_n(\theta) = E_n \phi_n(\theta),
 \end{equation}
with respect to boundary values determined by periodicity in
$\theta$ with period $2 \pi$. Evidently
 \begin{equation}\label{EVeq}
  E_n = \frac{n^2 \hbar^2}{2I},
   \qquad \phi_n(\theta) = e^{i n \theta}.
 \end{equation}
The solution of (\ref{SEeq}) satisfying the initial condition
 \begin{equation}\label{ICeq}
  \psi(\theta,0)= \dot{\delta} = \frac{1}{2\pi} \Sumii e^{i n (\theta-\theta_0)},
 \end{equation}
where $\dot{\delta}$ denotes the periodic delta distribution with
period $2\pi$ gives the kernel (or matrix element) of the
evolution operator $\exp -i H t/\hbar$:
 \begin{equation}\label{PKCeq}
  K(\theta,t;\theta_0,0)
  = \Dppp{\theta}{e^{\frac{-i H t}{\hbar}}}{\theta_0}
  = \frac{1}{2\pi} \Sumii e^{-\frac{i \hbar n^2 t}{2I}}
    e^{i n( \theta - \theta_0)},
 \end{equation}
which is convergent in the distribution sense; alternatively one
can give $t$ a small imaginary part to restore convergence. Later
we shall see that when both position and momentum are cyclic, with
phase space a torus, the problem of convergence does not arise.

From(\ref{PKCeq}) we see that the trace of the evolution operator
is
\begin{equation}\label{PTCeq}
  \Trace \left( \exp -i H t/\hbar \right)
  = \frac{1}{2\pi} \Sumii e^{-\frac{i \hbar n^2 t}{2I}}
\end{equation}
with convergence in the sense described above.

Now, as in \cite{Armita87}, we follow \cite{FeyHib} adapted to the
cylindrical phase space. The purpose of the section is to express
the old idea of using the method of images and the universal cover
of the circle in terms of path integrals, in such a way that the
argument can be adapted  to the toroidal phase space of section
\ref{TPSsec}, obtaining a second expression for the trace of the
evolution operator $\exp i H t/\hbar$ which on comparison with
(\ref{PTCeq}) gives the Jacobi identity.

Denote by
 \begin{equation}\label{ACCeq}
  S(\theta(t)) = \int_{0}^{t} \Lag(\dot{\theta}, \theta;t) dt,
 \end{equation}
the classical action, where $\Lag$ is the Lagrangian, and then we
obtain the evolution amplitude from $(\theta_0,0)$ to $(\theta,t)$
as
 \begin{equation}\label{PICeq}
   K(\theta,t;\theta_0,0)
    \sim \sum_{\mbox{periodic paths from }(\theta_0,0)\mbox{ to } (\theta,t)}
       e^{iS(\theta)/hbar}
 \end{equation}
where `$\sim$' indicates that a normalisation factor (c.f.
(\ref{PICAeq})) is required.  In order to carry out the path
integral sum we carry out the integration on the universal
covering space, $\Real$, of $S^1$, that is, we lift the homotopy
classes to the universal covering space \cite{Armita87,Schulm68}.
Corresponding to paths from $\theta_0$ to $\theta$ in $S^1$, we
have paths from some fixed $\theta_0^*$ in $\nu^{-1}(\theta_0)$
(here $\nu$ denotes the covering projection from $\Real$ to $S^1$)
to each of the $\theta_j^*$ in $\nu^{-1}(\theta)$, where the index
$j$ runs through the fundamental group ($\cong \Intg$) of $S^1$.
It follows (this is of course Kelvin's method of images) that
 \begin{equation}\label{PICAeq}
   K(\theta,t;\theta_0,0)
    = \Sumii \left( \frac{I}{2\pi i \hbar t}\right)^{\frac12}
     \exp\left( \frac{i I}{2 \hbar t} (\theta-\theta_0 - 2 \pi n)^2
     \right).
 \end{equation}
(The normalisation factor is taken from \cite{FeyHib}.)

On equating the expressions for the trace in (\ref{PTCeq}) and
(\ref{PICAeq}) we obtain
\begin{equation}\label{JIFeq}
  \Sumii e^{\frac{(i \hbar n^2 t)}{2I}}
 =  \left( \frac{2\pi I}{ i \hbar t}\right)^{\frac12} \Sumii
     e^\frac{-2\hbar^2 n^2 I}{ \hbar i t} .
\end{equation}
(As already observed, care is needed over convergence, which is
intended in the distributional sense.) If we arrange that $2\pi I
= \hbar$, $\tau =it$, then we obtain formally the usual Jacobi
identity (\ref{JIeq}) for $\vartheta(it)$.

After this work was presented \cite{Armita87} it was suggested by
Berry \cite{Berry1} that the way forward to obtaining the
Landsberg-Schaar formula by similar means might be to consider a
system where the angular momentum, as well as the angle, was
cyclic so that the phase space was compact and the quantized
observables would not only be discrete but also have a finite
range. This approach, which does indeed lead to the
Landsberg-Schaar formula, is developed in the following section.
\section{Toroidal phase space and the Landsberg-Schaar formula}
\label{TPSsec}
In  this section we adapt the methods of the preceding section to
a toroidal phase space on which both $\Amom$ and $\theta$ are
periodic.  On quantization this gives discreteness to both these
variables, and makes it possible (provided that suitable values
are used for the various constants involved) to have only a finite
number of energy eigenstates. Time, too, becomes quantized as
emerges when evolution is considered. In this section we
concentrate on ideas, obtaining the Landsberg-Schaar formula by
proceeding in the manner of a physicist, while in the next section
the quantization scheme is fully described and normalisation
factors are derived.

Suppose that the phase space of our system is a torus, with
angular momentum $\Amom$ of period $P$ and angle of rotation
$\theta$ of period $2\pi$. Quantization in $\Amom$ and $\theta$,
when both $\theta$ and $\Amom$ are periodic, gives, respectively:
\begin{equation}\label{PEReq}
  \Amom=n\hbar, \qquad \theta = \frac{2 \pi m\hbar}{P}, \qquad m,n \in \Intg.
\end{equation}
The phase space is quantized to a lattice on the torus, and the
small rectangles with sides $\hbar$ and $\frac{2 \pi \hbar}{P}$
fit a whole number of times into the phase space torus area. The
area of the phase space torus is $2\pi P$. So $2\pi P=2\pi
N\hbar$, ($P=N\hbar, N \in \Intg$) and so the small rectangles fit
$N^2$ times into the phase space torus.

There are precisely $N$ energies, $E_n,1 \leq n\leq N$, and so the
trace of the evolution operator is given by:
 \begin{eqnarray}\label{PTDeq}
 \Trace(\exp -iHt/\hbar) &=& \frac{1}{2\pi} \Sum{n=1}{N} e^{-\frac{iE_n t}{\hbar}} \End
   &=& \frac{1}{2\pi} \Sum{n=1}{N} e^{-\frac{i\hbar n^2 t}{2I}}.
 \end{eqnarray}
{\em Now we turn to the $\Amom$-periodicity and show that it
implies a restriction on the times, $t$, of `looking at the
system'.}

Any quantum mechanical state $\psi(\theta,t)$ has an expansion:
 \begin{equation}\label{EXeq}
  \psi(\theta,t) = \Sumkii a_k
  e^{i\left(k\theta-\frac{\hbar k^2t}{2I}\right)}, a_k=a_{k+N}
 \end{equation}
with the usual observations about convergence. Now take the
Fourier transform in $\theta$ of (\ref{EXeq}) to go from position
(angle) space to momentum space. We obtain (here $\Hphi$ denotes
the Fourier transform in the variable $\theta$),
 \begin{equation}
  \Hphi(\Amom,t) = \int_{-\infty}^{+\infty} \psi(\theta,t)
  e^{-ip\theta/\hbar}\, d\theta,
 \end{equation}
and we use ${\cal F}\left[e^{i k \theta} \right] =
\delta(\Amom-k\hbar)$)
 \begin{equation}
  \Hphi (\Amom,t) = \Sumkii a_k e^{-\frac{i\hbar k^2 t}{2 I}} \delta(\Amom-k\hbar),
 \end{equation}
where $\delta$ denotes the Dirac delta distribution, and
 \begin{eqnarray}
  \Hphi (\Amom,t)
  &=&   \Sumkii a_k e^{-\frac{i\hbar k^2 t}{2 I}} \delta(\Amom+P-k\hbar)\End
  &=&   \Sumkii a_k e^{-\frac{i\hbar (k+N)^2 t}{2 I}} \delta(\Amom-k\hbar).
 \end{eqnarray}
Thus periodicity of period $P$ in $\Amom$, that is
$\Hphi(\Amom+P,t) = \Hphi(\Amom,t)$, implies
 \begin{equation}
  1= e^{-i\hbar t(N^2+2kN)/(2I)}
 \end{equation}
 so that (assuming $N$ is even, as is required later)
 \begin{equation}\label{TQeq}
  t= \frac{2 \pi m I}{N \hbar}, \qquad m \in \Intg.
 \end{equation}
On substituting from (\ref{TQeq}) in (\ref{PTDeq}) we obtain
 \begin{eqnarray}\label{PTDTeq}
  \Trace(\exp -iHt/\hbar)
  &=& \frac{1}{2\pi} \Sum{n=0}N e^{-i\hbar n^2\frac{ 2\pi m I}{N \hbar 2I}}\End
   &=&  \frac{1}{2\pi} \Sum{n=0}N e^{-i\hbar\pi n^2\frac{m}{N}}.
 \end{eqnarray}

We have to compute
 \begin{equation}
  \sum_{\mbox{periodic paths}} e^{iS(\theta)/\hbar},
 \end{equation}
where the sum is now over all periodic paths on the torus. At the
$n^{th}$ energy level, $p_n$ is constant:
 \begin{equation}
  E_n = \frac{p_n^2}{2I}, I \dot{\theta}_n = p_n, I \theta = p_n t.
 \end{equation}
Hence, for the `time of looking' given in (\ref{TQeq}),
 \begin{equation}\label{LIMeq}
  |I\theta| \leq p_n . \frac{2|m| \pi I}{N \hbar} \leq |m|. 2 \pi I.
 \end{equation}
On repeating the argument given in section \ref{CPSsec}, but now
restricting $\theta$ as in (\ref{LIMeq}), we obtain in place of
(\ref{PICAeq})
\begin{equation}\label{PIDAeq}
   K(\theta,t)
    = \Sum{n=0}{m-1} \left( \frac{I}{2\pi i \hbar t}\right)^{\frac12}
     \exp\left( \frac{i I}{2 \hbar t} (\theta - 2 \pi n)^2 \right),
 \end{equation}
 where as before
 \begin{equation}
  t= \frac{2 \pi m I}{N \hbar}.
 \end{equation}
(The normalisation factor used is taken from (\ref{PICAeq}), and
is justified in section \ref{NORMsec}; while it might seem simpler
to derive this by the method used by Davison \cite{Daviso54} than
by the methods of section \ref{NORMsec}, in fact there is a
crucial step (equation (3.5) of Davison's  paper) which would
require knowledge of the Gauss sums under study.)

It follows from (\ref{PIDAeq}) that the propagator trace is
 \begin{eqnarray}\label{PTDAeq}
   \Trace (\exp - i H t /\hbar)
 &=&  \left( \frac{N}{4 \pi^2 i m} \right)^{\frac12} \Sum{n=0}{m-1}
    \exp \left(\frac{ i  n^2}{ \left(\frac{2 \hbar.2 m \pi}{N \hbar} \right)} \right) \End
 &=&   \left( \frac{N}{ i m} \right)^{\frac12}
   \Sum{n=0}{m-1} e^{i \pi n^2 N/m}.
\end{eqnarray}
So, on equating (\ref{PTDeq}) and (\ref{PTDAeq}), we obtain
 \begin{equation}\label{PENULMeq}
   \Sum{n=0}N e^{-\frac{ 2\pi i n^2}{N}}
  =  \left( \frac{N}{ i m} \right)^{\frac12}
   \Sum{n=0}{m-1} e^{i \pi n^2 N/m}.
 \end{equation}
Finally in (\ref{PENULMeq}) we take $N=2q$, $m=p$, $i^{-1}=-e^{\pi
i/2}$ to obtain
\begin{equation}
  \frac{e^{\pi i/4}}{\surd(2q)}
   \sum_{n=0}^{2q-1} \exp \left( - \frac{\pi i n^2 p}{2q}\right)
   =\frac{1}{\surd p} \sum_{n=0}^{p-1}
   \exp \left(  \frac{2\pi i n^2 q}{p} \right),
\end{equation}
which is the Landsberg-Schaar formula (\ref{LSeq}).
\section{Normalisation of the discrete path integral}
\label{NORMsec}
In this section  details are given of the quantum mechanical
system whose path integration is used in the previous section to
prove the Landsberg-Schaar formula (\ref{LSeq}). The necessary
path integral formula is derived by adapting the original approach
of Feynman \cite{FeyHib} to the discrete and cyclic setting.
Because all sums are finite in this case there are no convergence
or other analytic difficulties, so that a precise result is
obtained with well-defined normalisation.  This confirms  the
validity of the more heuristic, but geometrically and
arithmetically well-motivated, use of the Feynman principle in the
previous section.

The starting point is the classical phase space of the system,
which is taken to be toroidal with coordinates $\theta, \Amom$ of
periodicity $2\pi$ and $P$ respectively.  As Hilbert space for our
system we take $\Hil = \Comp^N$ (where $N$ is a positive integer)
realised as
 \[
  \left\{ f:\Real \to \Comp |
   f(\theta) = \Sum{n=1}{N} a_n \Exp{\frac{2\pi i n \theta}{N}}
   \right\}
  \]
where each $a_n$ is a complex number.

One basis of $\Hil$ is then plainly $\left\{f_k:k=0, \dots,N-1
\right\}$ with $f_k(\theta) = \frac{1}{\surd N} \exp (\frac{2\pi i
k \theta}{N})$. This is the angular momentum basis, that is, each
 $f_k$ is an eigenvector of the angular momentum operator $\Amom$
(defined now by
 $\Amom = -i \frac{N}{2\pi} \frac{\partial}{\partial\theta}$)
with eigenvalue $k$. Using Dirac notation, we write $f_k$ as
$\Ket{k}$. For simplicity we use units in which $\hbar$ takes the
value $1$.

Another basis is $\left\{b_r:r=1, \dots,N \right\}$ with
 \begin{equation}\label{POSBASeq}
  b_r(\theta) = \frac{1}{N}
  \Sum{k=0}{N-1} \exp\left(\frac{2\pi i(\theta-r)k}{N}\right).
 \end{equation}

The inner product on $\Hil$ is defined by
 \begin{equation}\label{IPDEFeq}
  \Dpp{f}{g} = \Sum{j=0}{N-1} \Cj{f(j)} g(j).
\end{equation}
With this inner product both of these  bases are orthonormal.

The $b_r$ may be regarded as the position basis if one restricts
the domain of the elements $f$ of $\Hil$ to
 $\Intg_N=\{0,1,\dots,N-1 \}$ and defines the
 (exponentiated angular) position
operator $\Apos$ by
 \begin{equation}\label{APOSeq}
  \Apos f(\theta) =\Exp{2 \pi i \frac{\theta}{N}}f(\theta),
 \end{equation}
since then
\begin{equation}\label{PEVeq}
  \Apos b_r =  \Exp{2 \pi i \frac{r}{N}} b_r
\end{equation}
so that $b_r$ is an eigenstate of $\Apos$, with eigenvalue
 $\Exp{2 \pi i \frac{r}{N}}  $.
(In Dirac notation we write $b_r$ as $\Ket{r}$.)

Now the proof of the Landsberg-Schaar identity essentially
involves calculating the trace of $\exp- i H t$ (with
 $H= \frac{\Amom^2}{2I}$ and $t= \frac{2 \pi m I}{N}$ as before)
in these two different bases. From the outset we will set $m=p$,
where $p$ is one of the two integers in the Landsberg-Schaar
formula.

{\em Method 1} is the direct way, that is, working in the momentum
basis in which $H$ is diagonal.

We have $H t \Ket{k} = \frac{\pi k^2 p}{N} \Ket{k}$ so that
  \begin{equation}\label{TR1eq}
  \Trace \left( \exp -i H t \right) =
   \Sum{k=0}{N-1} \Exp{-\frac{i \pi k^2 p}{N}}
 \end{equation}
\vskip 0.3cm
{\em Method 2} uses discrete, cyclic `path integrals'. We consider
 $\Dppp{r}{\exp -i H t}{s}$.

Breaking the time interval into $p$ steps $\Dt = \frac{t}{p}$, and
using $s_i$ to label basis elements $b_{s_i}$ at the $i^{th}$
step, we have
 \begin{eqnarray}
  \Dppp{r}{\exp -i H t}{s} &=& \Dppp{r}{\exp -i H \Dt}{s_{p-1}}
  \Dppp{s_{p-1}}{\exp -i H \Dt}{s_{p-2}} \End
  && \dots\Dppp{s_1}{\exp -i H \Dt}{s}
 \end{eqnarray}
(with summation from $0$ to $N-1$ over each of the intermediate
$s_i,i=1, \dots, p-1$). We thus need to consider
$\Dppp{s_{i}}{\exp -i H \Dt}{s_{i-1}}$.
Now
 \begin{eqnarray}
    \Dppp{s_{i}}{\exp -i H \Dt}{s_{i-1}}
    &=& \Sum{k=0}{N-1}\Dpp{s_i}{k} \Dpp{k}{s_{i-1}}
    \Exp{- \frac{\pi i k^2}{N}}   \End
  &=& \Sum{k=0}{N-1} \frac{1}{N} \Exp{\frac{2\pi i k(s_i-s_{i-1})}{N}}
 \Exp{- \frac{\pi i k^2}{N}}
 \end{eqnarray}
(using the fact that
 \begin{equation}
  \Dpp{r}{k} = \frac{1}{\surd N} \exp \frac{2 \pi k r}{N}).
 \end{equation}
Thus
 \begin{eqnarray}
    \Dppp{s_{i}}{\exp -i H \Dt}{s_{i-1}}
  &=&  \Sum{k=0}{N-1} \frac{1}{N}
  \Exp{-\frac{\pi i }{N}\left(k-(s_i-s_{i-1})\right)^2} \End
&& \quad  \Exp{\pi i\frac{ (s_i-s_{i-1})^2}{N}}.
 \end{eqnarray}

Again, as in section \ref{TPSsec}, we set $N=2q$, so that in
particular $N$ is even and we have
 \begin{eqnarray}
  \Dppp{s_{i}}{\exp -i H T}{s_{i-1}}
    &=&  \Sum{k=0}{N-1} \frac{1}{N}
  \Exp{-\frac{2\pi i }{2N}\left(k-(s_i-s_{i-1})\right)^2}
  \Exp{2\pi i \frac{(s_i-s_{i-1})^2}{2N}} \End
  &=& \frac{1}{\sqrt{i N}} \Exp{2\pi i \frac{(s_i-s_{i-1})^2}{2N}}
 \end{eqnarray}
by equation (\ref{AP2eq}) of the appendix. Thus
 \begin{equation}
  \Dppp{r}{\exp -i H T}{s} =(iN)^{-p/2}
  \sum_{s_1=0}^{N-1} \dots \sum_{s_{p-1}=0}^{N-1}
  \Exp{2\pi i \frac{\sum_{i=1}^p(s_i-s_{i-1})^2}{2N}}
 \end{equation}
where $s_0=r, s_p=s$.

At this stage we assume that $p$ and $q$ are coprime. Then, if $p$
is odd, so that  $p$ and $N=2q$ are coprime, we may observe that
 \begin{equation}\label{USEFULeq}
  \sum_{l=0}^{N-1} \sum_{k=0}^{p-1} \Exp{2\pi i \frac{(k N + l p)^2}{ 2N p}}
  = \sum_{t=0}^{N p-1} \exp \frac{2 \pi i t^2}{2N p} = \sqrt{ i N p}
 \end{equation}
where we have used the result (\ref{AP1eq}) from the appendix.
(Because $p$ and $N$ are coprime the expression $k N + l p$ takes
all $Np$ distinct values ($\Mod Np$) as $k$ ranges from $0$ to
$p-1$ and $l$ ranges from $0$ to $N-1$.) Also, if $p=2r$ is even
(but $p$ and $q$ are still coprime) then
 \begin{equation}
 \sum_{l=0}^{N-1}
 \sum_{k=0}^{p-1} \Exp{2\pi i \frac{(k N + l p)^2}{ 2N p}}
  =\sum_{l=0}^{2q-1}\sum_{k=0}^{2r-1}
  \Exp{2\pi i \frac{(k q + l r)^2}{ 2qr}}
  = \sqrt{(1+(-1)^r) i N p}.
 \end{equation}
This result allows us to insert in the expression for
$\Dppp{r}{\exp -i H t}{s}$ the extra summations $\sum_{l=0}^{N-1}
 \sum_{k=0}^{p-1} \Exp{2\pi i \frac{(k N + l p)^2}{ 2N p}}$
together with the compensating factor $1/\sqrt{ i N p}$ (if $p$ is
odd) or $1/\sqrt{ 2i N p}$ (if $p=2r$ with $r=2r'$ even). We can
overlook the case where $p=2r$ and $r$ is odd, since in that case
the Landsberg-Schaar equation is trivially satisfied. This step
effectively allows the winding number $k$ for a path to be shared
over the $p$ steps in the path.

Using these facts, we see that if $p$ and $q$ are coprime and
$a=1$ when $p$ is odd while $a=2$ when $p=4r'$,
 \begin{eqnarray}
   \lefteqn{\Dppp{r}{\exp -i H t}{s}  } \End
   &=&\Factx
   \sum_{s_1=0}^{N-1} \dots \sum_{s_{p-1}=0}^{N-1}
   \sum_{l=0}^{N-1} \sum_{k=0}^{p-1}
  \Exp{2\pi i \left(\sum_{i=1}^p\frac{(s_i-s_{i-1})^2}{2N}
  +  \frac{(k N + l p)^2}{ 2 N p }\right)}\End
  &=&  \Factx
   \sum_{s_1=0}^{N-1} \dots \sum_{s_{p-1}=0}^{N-1}
   \sum_{l=0}^{N-1} \sum_{k=0}^{p-1}
  \Exp{2\pi i \left(\sum_{i=1}^p
  \left(\frac{(s_i-s_{i-1})^2}{2N}+\frac{l^2}{2N}\right)
  +  \frac{(k^2 q)}{ p }\right)}\End
  &=& \Factx
   \sum_{s_1=0}^{N-1} \dots \sum_{s_{p-1}=0}^{N-1}
   \sum_{l=0}^{N-1} \sum_{k=0}^{p-1}
  \Exp{2\pi i \left(\sum_{i=1}^p\frac{\left(s_i-s_{i-1}+l\right)^2}{2N}
  +  \frac{(k^2 q)}{ p }\right)}\End
  \end{eqnarray}
provided that $r-s\equiv 0 \,(\Mod \, N)$.

Now let $u_1=s_1-s_0+l=s_1-r+l$, $u_2=s_2-s_1+l$ and so on, with
  $u_p=s_p-s_{p-1}+l  =s-s_{p-1}+l$. Then $\sum_{i=1}^p u_i=r-s+lp$ so that
$u_p=r-s+lp - \sum_{i=1}^{p-1} i_i$. Thus if $p$ and $N$ are
coprime and $s_1, \dots, s_{p-1}$ are fixed  the integer variable
$u_p$ will take each value $0,1,\dots,N-1 \, \Mod\,N$ precisely
once as $l$ ranges from $0$ to $N-1$, so that
  \begin{eqnarray}
   \lefteqn{\Dppp{r}{\exp -i H T}{s}  } \End
  &=&  \Facty
   \sum_{u_1=0}^{N-1} \dots \sum_{u_{p}=0}^{N-1}
    \sum_{k=0}^{p-1}
  \Exp{2\pi i \left(\sum_{i=1}^{p}\frac{u_i^2}{2N}\right)}
   \Exp{2\pi i  \frac{(k^2 q)}{ p }}\End
  &=&  \Facty (N i)^{p/2} \sum_{k=0}^{p-1}
  \Exp{2\pi i  \frac{(k^2 q)}{ p } }\mbox{by (\ref{AP1eq})} \End
  &=& \frac{1}{\sqrt{i p N}} \sum_{k=0}^{p-1}
  \Exp{2\pi i  \frac{(k^2 q)}{ p }},
 \end{eqnarray}

Hence
 \begin{eqnarray}
  \Trace \left(\exp i H t \right)
  &=& \Sum{s=0}{N-1} \Dppp{s}{\exp i H t}{s}   \End
  &=& \frac{N}{\sqrt{i p N}}\sum_{k=1}^p
  \Exp{2\pi i  \frac{(k^2 q)}{ p }} \End
  &=& \frac{\sqrt{2q}}{\sqrt{i p }}\sum_{k=1}^p
  \Exp{2\pi i  \frac{(k^2 q)}{ p }} \qquad (\mbox{since\ }N=2q).
 \end{eqnarray}
(If $p=4r'$ with $p$ and $q$ still coprime, then the values of
$u_p$ are restricted, so that summation over $l$ contributes a
factor $ \Exp{\frac{u_i{}^2}{2N}} = \sqrt{2 i N p}$. The result
above is thus also obtained in this case.)

Hence in all cases when $p$ and $q$ are coprime we have (using
(\ref{TR1eq}),
\begin{equation}
   \Sum{k=0}{2q-1} \Exp{- \pi i  \frac{ k^2 p}{2q}}
   =
   \frac{\sqrt{2q}}{\sqrt{i p }}\sum_{k=0}^{p-1}
  \Exp{2\pi i  \frac{(k^2 q)}{ p }}
\end{equation}
which immediately gives the Landsberg-Schaar formula when $p$ and
$q$ are coprime. The general case follows on setting $p=mp'$,
$q=mq'$ with $p',q'$ coprime.
\vskip 0.3cm
 \noindent {\Large \bf Appendix}
  \vskip 0.2cm \noindent
The two formulae below may be proved by elementary means:
\cite{Landau27,Waterh70}
  \begin{equation}\label{AP1eq}
  \Sum{n=0}{2r-1} \Exp{ \frac{2\pi i (n-s)^2}{4r}}= \sqrt{2ri}
 \end{equation}
 \begin{equation}\label{AP2eq}
  \Sum{k=0}{2r-1} \Exp{-2\pi i \frac{(k-s)^2}{4r}}= \sqrt{\frac{2r}{i}}
 \end{equation}
Given such an elementary evaluation of the Gauss sum, together
with an elementary proof of the quadratic reciprocity law
 \begin{equation}
 \left( \frac{p}{q} \right) \left( \frac{p}{q} \right) =
   (-1)^{\frac{p-1}{2} \, \frac{q-1}{2}},
 \end{equation}
where $\left( \frac{p}{q} \right)$ denotes the Legendre symbol and
$p$ and $q$ are odd primes, one could prove (\ref{LSeq}) by
evaluating each side and then appealing to the reciprocity law,
and its extensions (to $p=2$ and $p=-1$). Whether such a proof
would provide insight into why the result is true is another
matter.

Again, there is another proof of the theta function identity
(\ref{JIeq}), due to Polya \cite{Polya27}, which depends on an
identity involving binomial coefficients, identities which are in
turn related to the Markov chain approach to diffusion processes
considered in \cite{Armita87}.

The approach adopted in this paper presupposes an elementary proof
of the evaluation of Gauss sums in (\ref{AP1eq}) and
(\ref{AP2eq}), but not of the reciprocity law. It is perhaps
tempting therefore to regard the present approach as a substitute,
in some sense, to the reciprocity law, but we prefer to see it as
a quantum mechanical equivalent of Hecke's observation, in which
Feynman's `sum over histories' in discrete time replaces the
limiting process that derives (\ref{LSeq}) from (\ref{JIeq}).
 %

\end{document}